\documentclass[prl,preprint,aps]{revtex4}
\usepackage{graphicx}

\newcommand{\Rb}{^{87}\mathrm{Rb}}
\newcommand{\kB}{k_{\mathrm{B}}}
\renewcommand{\vec}[1]{{\mathbf{#1}}}
\newcommand{\etal}{\emph{et~al.}}

\date{\today}

\begin{document}

\title{Understanding the production of dual BEC with sympathetic cooling}

\author{G. Delannoy\footnote{guillaume.delannoy@iota.u-psud.fr, http://atomoptic.iota.u-psud.fr},
S. G. Murdoch, V. Boyer, V. Josse, P. Bouyer and A. Aspect}

\affiliation{Groupe d'Optique Atomique Laboratoire Charles Fabry
de l'Institut d'Optique, UMRA 8501 du CNRS, B\^{a}t. 503, Campus
universitaire d'Orsay, B.P. 147, F-91403 ORSAY CEDEX, FRANCE}

\begin{abstract}
We show, both experimentally and theoretically, that sympathetic
cooling of $^{87}$Rb atoms in the $|F=2,m_F=2\rangle$ state by
evaporatively cooled atoms in the $|F=1,m_F=-1\rangle$ state can
be precisely controlled to produce dual or single condensate in
either state. We also study the thermalization rate between two
species. Our model renders a quantitative account of the observed
role of the overlap between the two clouds and points out that
sympathetic cooling becomes inefficient when the masses are very
different. Our calculation also yields an analytical expression of
the thermalization rate for a single species.
\end{abstract}


\maketitle

Evaporative cooling has proved to be an efficient route toward
Bose-Einstein condensation (BEC). However, not all the species are
eligible for evaporation. For instance, the collisions may not be
sufficient \cite{weiman} or may be forbidden as for fermions at
low temperature \cite{jin}. In the case of a rare species, one may
also want to avoid the large loss of atoms inherent to evaporative
cooling. Sympathetic cooling is potentially a very good solution
in those cases. It consists in putting the target sample we want
to cool in thermal contact with a buffer gas that can be cooled to
the desired temperature. This cooling can be done either by
conventional cryogenics \cite{doyle,Wineland}, or by evaporative
cooling as demonstrated by Myatt \etal \, \cite{cornell}. These
authors have produced dual BEC by sympathetic cooling of a gas of
$^{87}$Rb atoms in the $|F=2,m_F=2\rangle$ state (target gas,
noted $F=2$ hereafter) by the use of an evaporatively cooled gas
of $^{87}$Rb atoms in the $|F=1,m_F=-1\rangle$ state (buffer gas
noted $F=1$). In their experiment, the number of target atoms,
although almost constant during the initial phase of the cooling,
decreases significantly near to the condensation. In this paper,
we investigate a situation similar to the one of Myatt \etal, but
with the number of atoms in the target gas constant in the last
cooling stage, all the way down to BEC. This allows us to study
quantitatively how the production of dual BEC relies on a precise
control of the initial conditions. We also study the effect of the
interspecies thermalization rate on sympathetic cooling
efficiency, by changing the overlap between the buffer and the
target gas. Our experimental observations agrees with the results
of a calculation of the thermal contact between the atomic
species. This calculation not only shows the crucial role of the
overlap, but also why sympathetic cooling is very efficient for
equal masses of the target and buffer gases. In addition, it
allows us to recover in an analytic way a result previously found
numerically for the thermalization rate of a single species.

Our experimental apparatus has been described in a previous
publication \cite{bruno}. After a laser cooling sequence, we
transfer the $^{87}$Rb atoms from a dark SPOT into the magnetic
trap. During the transfer, an optical pumping pulse allows us to
transfer most of the atoms in $F=1$ (buffer), while keeping an
adjustable fraction in $F=2$ (target). Our iron-core
Ioffe-Pritchard trap operates with a high bias field $B_{0}$
between 50 and 250\,Gauss, and the magnetic field curvature $C$
along the dipole direction is linked to $B_0$ (the ratio $B_{0}/C$
is fixed by construction to 1\,cm$^{2}$). After adiabatic
compression, the magnetic field gradient in the strong
(quadrupole) directions is $G = 1\,$kG/cm. To cool down $F=1$, we
use standard rf forced evaporation. We characterize both species
simultaneously by absorption imaging after turning off the
magnetic trap. During this release, the atoms experience a
magnetic kick due to a Stern-Gerlach effect which allows us to
spatially separate atoms in $F=1$ and atoms in $F=2$. We probe on
the $F=2 \rightarrow F^{\prime} =3$ transition after repumping the
atoms from $F=1$ to $F=2$. With usual analysis, we derive the
temperature of the two clouds from the image.

In a first experiment, we have studied sympathetic cooling at a
bias of 56\,G, with initially $10^8$ atoms in $F=1$ at a
temperature of 300\,$\mu$K, and various initial numbers of atoms
in $F=2$. We observe no loss of target atoms during the whole
cooling process \cite{inelas}. Indeed, the atoms in $F=2$ see a
trapping potential twice as steep as for $F=1$, and as in
\cite{cornell}, very few target atoms are evaporated at the
beginning of the evaporation sequence. Moreover, at our high bias
field, the non linear Zeeman effect prevents rf coupling of the
$|F=2,m_F=2\rangle$ trapping state to any non trapping state at
the end of the evaporation \cite{bruno,3cou}. When starting with a
very small ($\leq 3\times 10^4$) number of atoms in $F=2$, we get
a $F=1$ condensate with typically $10^6$ atoms. If we continue the
evaporation, another condensate appears in $F=2$ at a lower
temperature. For $3\times 10^4$ atoms in $F=2$, both gas condense
simultaneously at a temperature of 200\,nK, as shown figure
\ref{evap}.a. If we start with a larger number of target atoms
(but smaller than $8 \times 10^4$), a condensate appears first in
$F=2$ at $T > 200$\,nK, and then in $F=1$ at $T < 200$\,nK. For a
yet larger number of target atoms, we can obtain a condensate in
$F=2$ and no condensate in $F=1$. If the number of atoms in $F=2$
is more than $2 \times 10^5$, BEC is observed in neither hyperfine
state.

These experimental observations can be explained with a simple
model based on an energy budget. We assume that both species are
always thermalized (see last part of this paper), and that the
number $N_2$ of target atoms is constant. The total energy of the
$N_1$ buffer and $N_2$ target atoms in harmonic traps is therefore
$E = 3 (N_1+N_2) \kB T$. The buffer gas is evaporatively cooled
with an energy cutoff $\eta \, \kB T$. We denote $dN_1$ the
elementary (negative) variation of the number of buffer atoms. The
corresponding (negative) energy variation is $dE= dN_1 (\eta + 1)
\kB T$ \cite{luiten}. After rethermalization at a lower
temperature $T+dT$, the total energy becomes $E + dE = 3(N_1 +
dN_1 + N_2) \kB (T + dT)$. Keeping only the first order, we have
$dT /T = \alpha \, dN_1 / (N_1 + N_2)$ with $\alpha = (\eta - 2)
/3$. Assuming that $\eta$ is constant, we get:

\begin{equation}
T = T^{\rm min} \left( \frac{N_1} {N_2} + 1 \right) ^\alpha \:
\textrm{with} \: T^{\rm min} = T^{\rm ini} \left( \frac {N_2}
{N_1^{\rm ini}} \right) ^\alpha \! \! . \label{sympat}
\end{equation}
In this formula, $T^{\rm ini}$ and $N_1^{\rm ini}$ are the initial
values, $N_2$ is constant and has been neglected with respect to
$N_1^{\rm ini}$ in the expression of $T^{\rm min}$ (as the
temperature has to be reduced by several orders of magnitude, the
initial fraction of atoms in the target gas must be very small).
The temperature $T^{\rm min}$ is reached when all the buffer atoms
have been evaporated ($N_1=0$). To understand the different
regimes experimentally observed, we first study the evolution of
the phase space densities ${\mathcal D}_1$ of the buffer and
${\mathcal D}_2$ of the target gas, in harmonic traps of mean
frequencies $\omega_1$ and $\omega_2$, during the evaporation.
Equation \ref{sympat} allows us to express ${\mathcal D}_1$ and
${\mathcal D}_2$ as functions of $N_1$ only. Figure \ref{courbe}
represents the evolution of ${\mathcal D}_1$ and ${\mathcal D}_2$
during the evaporation (while $N_1$ is decreasing). We observe
that the phase space density of the target gas ${\mathcal D}_2$
always increases to reach its maximum value

\begin{equation}
{\mathcal D}_2^{\rm max} = \frac {2.17} {N_2^{(3\alpha-1)}} \left(
{N_1^{\rm ini}}^\alpha \frac{ \hbar {\omega}_2}  {\kB T^{\rm ini}}
\right)^3 \label{target}
\end{equation}
at the very end of the evaporation \cite{num}; this is because
$N_2$ is constant and the temperature always decreases. We also
observe that the buffer gas phase space density ${\mathcal D}_1$
first increases \cite{d1} to a maximum ${\mathcal D}_1^{\rm max}$
that we write

\begin{equation} {\mathcal D}_1^{\rm max} =
{\mathcal D}_2^{\rm max} \: \left( \frac {{\omega}_1} {{\omega}_2}
\right) ^{\! 3} \: \frac {(3 \alpha - 1) ^{(3 \alpha -1)}} {(3
\alpha) ^ {3 \alpha}}, \label{buffer}
\end{equation}
and then decreases to zero. At the intersection of the two curves,
the phase space densities ${\mathcal D}_1$ and ${\mathcal D}_2$
have the same value
\begin{equation}
{\mathcal D}^{(=)} = {\mathcal D}_2^{\rm max} \: \left( 1+ \left(
\frac {{\omega}_2} {{\omega}_1} \right) ^3 \right) ^{-3 \alpha}.
\label{first}
\end{equation}
We note that a modification of the various parameters only changes
the relative position of the two curves, in particular the
ordering  of ${\mathcal D}^{(=)}$, ${\mathcal D}_1^{\rm max}$ and
${\mathcal D}_2^{\rm max}$. This is what determines the outcome of
sympathetic cooling.

We specifically focus on the situation where ${\mathcal D}^{(=)}$
is reached before ${\mathcal D}_1^{\rm max}$ (this happens if
$3\alpha - 1 > ({ \omega}_1 / { \omega}_2 )^3$), which corresponds
to the case of figure \ref{courbe} and to our experimental
parameters. The target gas can condense only if ${\mathcal
D}_2^{\rm max} > 2.612$, and in the same way, the buffer gas can
condense only if ${\mathcal D}_1^{\rm max} > 2.612$. If these two
conditions are satisfied, the buffer condenses first if ${\mathcal
D}^{(=)} > 2.612$ otherwise the target condenses first. These
three conditions can be written as conditions on the number $N_2$
of target atoms. This leads to define three critical numbers
$N_2^a$, $N_2^b$ and $N_2^c$ of target atoms, for which ${\mathcal
D}^{(=)}$, ${\mathcal D}_1^{\rm max}$ and ${\mathcal D}_2^{\rm
max}$ are respectively equal to 2.612. Although each of these
numbers depend on all initial conditions, the ratios $N_2^a/N_2^c$
and $N_2^b/N_2^c$ are independent of the initial values $N_1^{\rm
ini}$ and $T^{\rm ini}$ because ${\mathcal D}_1^{\rm
max}/{\mathcal D}_2^{\rm max}$ and ${\mathcal D}^{(=)}/{\mathcal
D}_2^{\rm max}$ do not depend on the initial conditions (see eq.
\ref{buffer} and \ref{first}) and because ${\mathcal D}_2^{\rm
max} \propto N_2 ^ {(1 - 3 \alpha)}$ (see eq. \ref{target}). These
ratios are represented in a diagram of the energy cutoff $\eta = 3
\alpha + 2$ versus the number of target atoms $N_2/N_2^c$ on
figure \ref{res}. We can identify four regions on this diagram
\cite{foot3}, corresponding to the different regimes of
sympathetic cooling. From the experiment, we measure $N_2^a= 3
\times 10^4$, $N_2^b= 8 \times 10^4$ and $N_2^c= 2 \times 10^5$
which is in good agreement with the calculated values for a
typical value of the evaporation parameter $\eta \simeq 6.5$ (see
fig. \ref{res}).

In a second experiment, we have studied the role of thermalization
in sympathetic cooling by changing the bias field from  56\,G to
207\,G. This reduces the vertical oscillation frequencies
$\omega_{F z} = \{ (-1)^F m_F \, \mu_{\rm B} (G^2 / B_0 - C) / 2 M
\}^{1/2}$ where $M$ is the mass of the atom and $\mu_{\rm B}$ the
Bohr magneton, and therefore increases the relative gravitational
sag $\Delta = g/ \omega_{1 z} ^{\, 2} - g/ \omega_{2 z} ^{\, 2}$
between the two atomic clouds \cite{cornell}. Figure \ref{evap}.b
represent successive absorption images while decreasing the final
frequency of the evaporation ramp, at a bias field of 207\,G
($\Delta=26\,\mu$m). We initially observe sympathetic cooling of
the target gas until a temperature of about 400\,nK is reached.
Then sympathetic cooling stops. This is in contrast with the first
experiment at a bias field of 56\,G, where the relative sag is
smaller ($\Delta=7\,\mu$m), and sympathetic cooling works all the
way down to BEC of the two species (see fig. \ref{evap}.a).

To render a quantitative account of these observations, we study
how the thermal contact between the two clouds evolves during the
cooling. For that, we calculate the energy exchange rate $W$ via
interspecies elastic collisions. We assume Maxwell-Boltzman
distribution at temperatures $T_1$ for the buffer and $T_2$ for
the target gas (we neglect the cut-off due to possible
evaporation). Introducing the notations $\vec{v_G}$ for the center
of mass velocity, $\vec{v}$ and $\vec{v}^\prime$ for the relative
velocities respectively before and after the collision, the energy
received in the laboratory frame by a buffer atom during the
collision is $\frac{M}{2} \, \vec{v_G} \cdot (\vec{v} -
\vec{v}^\prime )$. At low temperature we only consider $s$-wave
scattering and the term $\vec{v_G} \cdot \vec{v}^{\prime}$
averages out to zero. After integration over positions and
velocities, we get $W = \kB (T_2-T_1) \, \Gamma$, where $\Gamma$
is the number of interspecies collisions per unit of time, found
equal to

\begin{equation}
\Gamma \, =  \frac {N_1 \, N_2} {\pi^2 \, \rho_x \rho_y \rho_z}
\times \sigma_{12} \times V \times \exp \! \left( - \frac
{\Delta^2} {2 \, \rho_z^2} \right). \label{Gamma}
\end{equation}
In equation \ref{Gamma}, $\sigma_{12}$ is the interspecies elastic
cross-section (taken independent of the temperature), $V =
\sqrt{\kB (T_1 + T_2) /M}$ is the RMS sum of thermal velocities,
and $\rho_z = \sqrt{ \frac {\kB} {M} ( \frac {T_1} {\omega_{1
z}^{\, 2}} + \frac {T_2} {\omega_{2 z}^{\, 2}})}$ is the RMS sum
of the vertical sizes of the clouds (with similar expressions for
$\rho_x$ and $\rho_y$). The exponential term in (\ref{Gamma})
describes the overlap of the two clouds. Thus the energy exchange
rate $W$ will become vanishingly small if $\rho_z$ becomes smaller
than the gravitational sag $\Delta$. We can now analyze the role
of the thermal contact in the experiments. At a bias field of
207\,G (sag $\Delta=26\,\mu$m), and high temperature (above
400\,nK), the effect of the sag is negligible and the clouds are
thermalized. During the cooling, the sizes of the clouds and
therefore $\rho_z$ decreases. At the temperature of $T_1 \simeq
T_2 \simeq 400\,$nK where we observed that sympathetic cooling
stops, we have $\rho_z= 12\,\mu$m and the exponential term in
equation \ref{Gamma} takes the value 0.1, so that the energy
exchange rate is reduced by an order of magnitude. In contrast, at
a bias field of 56\,G, where $\Delta=7\,\mu$m, the parameter
$\rho_z$ at the temperature of condensation 300\,nK,
($\rho_z\simeq 8\,\mu$m) is still bigger than $\Delta$, the
collision rate $\Gamma$ is only reduced by 30 \% because of the
overlap, and sympathetic cooling works until reaching BEC.

Starting from the interspecies energy exchange rate $W$, and using
the total energy conservation, we can derive evolution equations
for the temperatures $T_1$ and $T_2$ of each species taken
separately at thermal equilibrium. Denoting $\Delta T= T_1-T_2$,
and $T=(T_1+T_2)/2$, we obtain an interspecies thermalization
rate, which, for equal trapping frequencies ($\omega_1 =\omega_2
=\omega$), takes the simple form
\begin{equation}
\frac {1} {\tau} = - \frac {1} {\Delta T} \frac {d\,\Delta T} {dt}
= \frac {(N_1 + N_2)} {3 \kB T} \frac {{\omega}^3 \, \sigma_{12}
\, M} {2 \, \pi^2}. \label{time}
\end{equation}
This result is remarkable, by analogy with the individual
thermalization rate of a single species $\tau^{-1} \simeq \gamma
/3$ \cite{simul}, where $\gamma = N \omega^3 \sigma M / 2\pi^2 \kB
T$ is the average elastic collision rate in an harmonic trap.
Assuming identical values for the elastic cross sections
$\sigma_1$, $\sigma_2$, and $\sigma_{12}$ \cite{Rb}, we then have
$\tau^{-1} \simeq \tau_1^{-1} + \tau_2^{-1}$ where $\tau_1^{-1}$
and $\tau_2^{-1}$ are the thermalization rates of each species
considered separately. This means that thermalization between the
two species happens faster than thermalization of each species.
So, if the evaporation ramp is adapted to cooling of the buffer
alone, interspecies thermalization will be efficient.

If the masses are different, equation \ref{time} remains valid at
lower order in $\Delta T/T$, if we replace $M$ by ${\mathcal M} =
\frac {8 (M_1M_2) ^2} {(M_1+M_2) ^3}$, which is symmetric in the
masses $M_1$ of the buffer and $M_2$ of the target. The equivalent
mass ${\cal M}$ is maximum for $M_1=M_2$ as in our experiment, and
decreases for uneven masses (for significant different masses,
${\cal M}$ is much smaller than the smallest mass). It is
interesting to note that if equation \ref{time} is applied to any
arbitrary partition in a single species, where $N_1$ and
$N_2=N-N_1$ atoms are taken at different temperatures, then the
thermalization rate between the two subsamples is exactly equal to
$\gamma/3$, in agreement with the numerical simulations of
\cite{simul}. To our knowledge, such an analytical result had not
been published.

To conclude, we have first presented a simple analysis allowing
one to quantitatively predict the outcome of sympathetic cooling,
in the case of no losses in the target. We have then obtained an
expression of the thermalization rate between the buffer and the
target, that quantitatively describes the effect of the overlap
between the two clouds. Moreover, when the overlap is good, the
interspecies thermalization rate is the same of the intraspecies
thermalization rate, and sympathetic cooling is very efficient
\cite{jin,Salomon}. In contrast, we predict a strongly diminished
interspecies rethermalization rate for very different masses of
the buffer and target. Our approach allows to derive an analytical
expression of the intraspecies thermalization rate in surprisingly
good agreement with previous numerical calculation. Although our
modeling is very simple and does not take into account quantum
statistics effects near condensation \cite{deg}, it is a useful
tool, that can easily be generalized, to plan or analyze
experiments using sympathetic cooling.

\begin{acknowledgments}
The authors thank M. L\'ecrivain for his contribution to the
experimental set-up. This work is supported by CNRS, MENRT,
R\'{e}gion Ile de France, DGA and the European Community. S.G.M.
acknowledges support from Minist\`{e}re des Affaires
\'{E}trang\`{e}res.
\end{acknowledgments}

\begin{figure}
\caption{Sympathetic cooling of $\Rb$ atoms in $|F=2,m_F=2\rangle$
(upper images) by evaporatively cooled atoms in $|F=1,
m_F=-1\rangle$ (lower images): effect of the thermal contact. {\bf
a)} Small sag ($7\,\mu$m) between the two clouds: sympathetic
cooling yields BEC for both species. {\bf b)} Larger sag
($26\,\mu$m): sympathetic cooling stops after the third snapshot
(400\,nK) when decreasing final rf frequency.} \label{evap}
\end{figure}

\begin{figure}
\caption{Phase space densities of the buffer (solid line) and the
target gas (dashed) as a function of the evaporation progress,
characterized by the ratio $N_1/N_1^{\rm ini}$ of remaining buffer
atoms. This plot corresponds to the case $3\alpha - 1 >
({\omega}_1 / {\omega}_2)^3$. The horizontal line is the critical
value 2.612 above which BEC occurs. Here, the target and then the
buffer will condense.} \label{courbe}
\end{figure}

\begin{figure}
\caption{Outcome of sympathetic cooling as a function of the
number of target atoms $N_{2}/N_2^{c}$. The evaporation parameter
$\eta$ is the ratio of the energy cutoff to the thermal energy
$\kB T$. Trap parameters are fixed to ${\omega}_{2} / {
\omega}_{1} = \sqrt{2}$.  The lines represents the critical number
ratios $N_2^{a}/N_2^{c}$, $N_2^{b}/N_2^{c}$ and $N_2^{c}/N_2^{c}$.
They separate regions where either dual, single or no BEC can be
formed. Black dots indicate the critical number found in our
experiment.} \label{res}
\end{figure}


\begin{thebibliography}{2}
\bibitem{inelas}
In contrast to the observations of ref \cite{cornell}, there are
no inelastic collisions in our situation.
\bibitem{num}
The factor of 2.17, related to the quantum statistics effect, is
valid close to degeneracy.
\bibitem{d1}
This is true for $3 \alpha > 1$, which is the usual condition for
the phase space density to increase during evaporation of a single
specie.
\bibitem{foot3}
In addition, there is a small region corresponding to $(
{\omega}_1 / {\omega}_2 )^3 > 3 \alpha - 1 > 0$, where a
condensate appears in the buffer gas, then desapears and finally
the target gas reaches degeneracy. This case is relevant if
${\omega}_1 > {\omega}_2$.
\end{thebibliography}
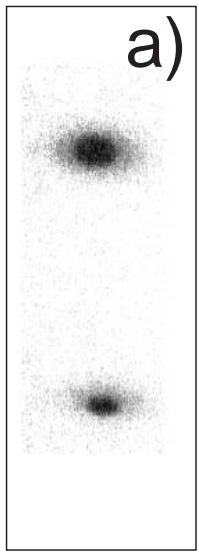
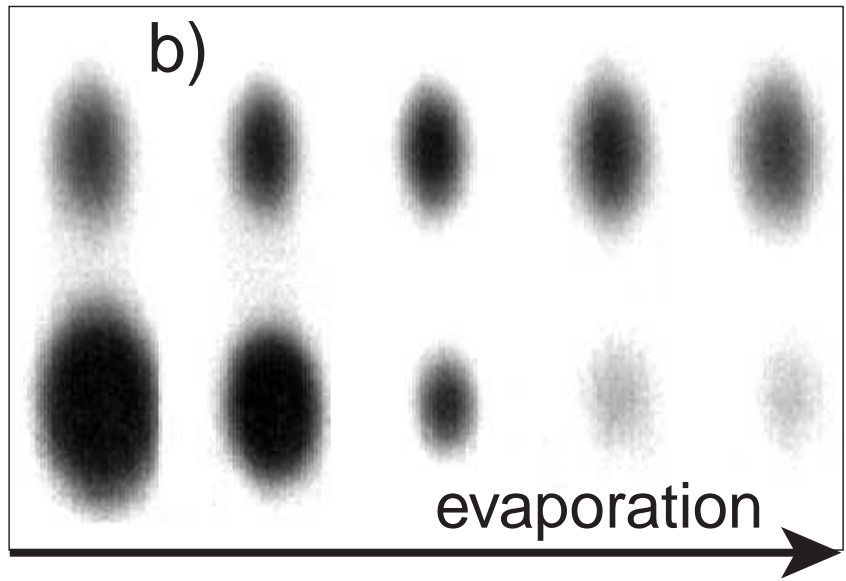
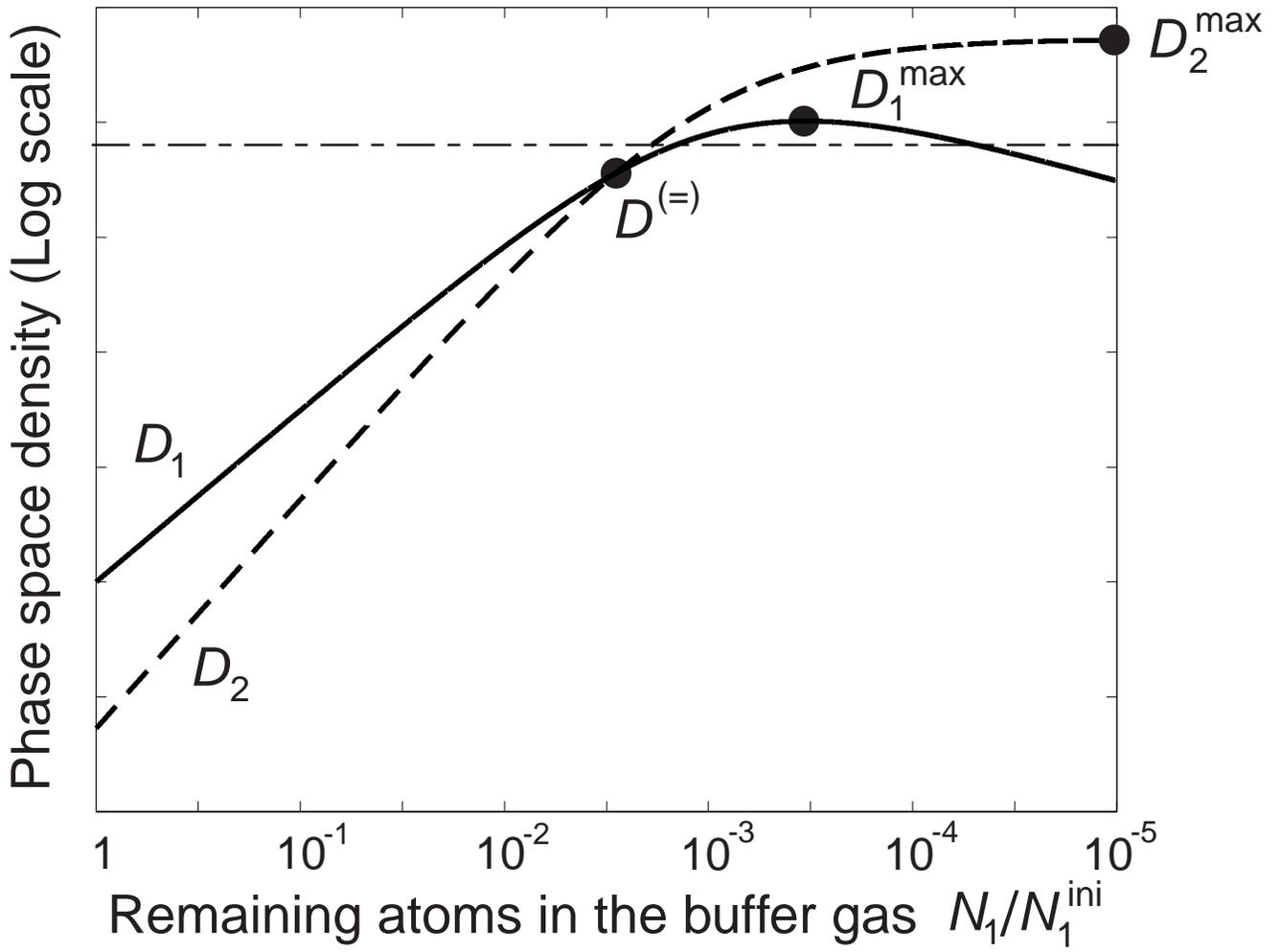
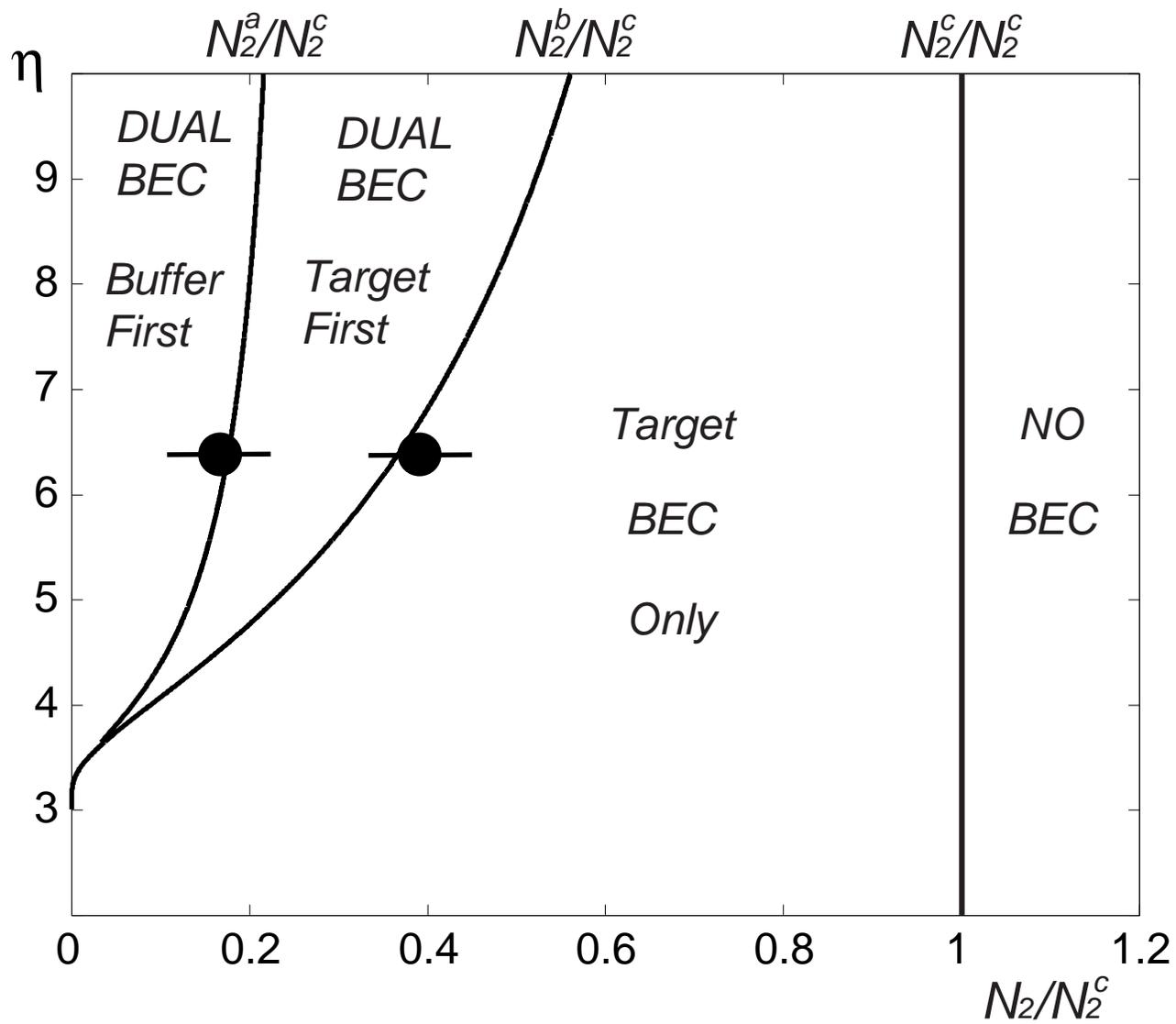
\end{document}